\documentclass{emulateapj}
\usepackage{natbib}
\shorttitle{OBSERVED LIMITS ON CHARGE EXCHANGE}\shortauthors{CROWDER ET AL.}

\begin{document}

\title{Observed Limits on Charge Exchange Contributions to the Diffuse X-ray Background}
\author{S. G. Crowder\altaffilmark{1,2}, K. A. Barger\altaffilmark{1}, D. E. Brandl\altaffilmark{1}, M. E. Eckart\altaffilmark{3,4}, M. Galeazzi\altaffilmark{5}, R. L. Kelley\altaffilmark{3}, C. A. Kilbourne\altaffilmark{3}, D. McCammon\altaffilmark{1}, C. G. Pfendner\altaffilmark{1}, F. S. Porter\altaffilmark{3}, L. Rocks\altaffilmark{1,6}, A. E. Szymkowiak\altaffilmark{7}, and I. M. Teplin\altaffilmark{1}}
\affil{\altaffilmark{1}Department of Physics, University of Wisconsin, Madison, WI 53706, USA\\
\altaffilmark{2}present address: School of Physics and Astronomy, University of Minnesota, Minneapolis, MN 55455, USA; crowder@physics.umn.edu\\
\altaffilmark{3}NASA Goddard Space Flight Center, Greenbelt, MD 20771, USA\\
\altaffilmark{4}CRESST and University of Maryland, Baltimore County, MD 21250, USA\\
\altaffilmark{5}Department of Physics, University of Miami, Coral Gables, FL 33124, USA\\
\altaffilmark{6}present address: Department of Science and Technology, Front Range Community College, Westminster, CO 80031, USA\\
\altaffilmark{7}Department of Physics, Yale University, New Haven, CT 06511, USA}

\begin{abstract}
We present a high resolution spectrum of the diffuse X$\mbox{-}$ray background from 0.1 to 1~keV for a $\sim1~{\rm sr}$ region of the sky centered at $l=90\degr{}$, $b=+60\degr{}$ using a 36$\mbox{-}$pixel array of microcalorimeters flown on a sounding rocket.  With an energy resolution of 11~eV~FWHM below 1~keV, the spectrum's observed line ratios help separate charge exchange contributions originating within the heliosphere from thermal emission of hot gas in the interstellar medium.  The X$\mbox{-}$ray sensitivity below 1~keV was reduced by about a factor of four from contamination that occurred early in the flight, limiting the significance of the results.  The observed centroid of helium$\mbox{-}$like \ion{O}{7} is $568^{+2}_{-3}~{\rm eV}$ at 90\% confidence.  Since the centroid expected for thermal emission is 568.4~eV while for charge exchange is 564.2~eV, thermal emission appears to dominate for this line complex, consistent with much of the high$\mbox{-}$latitude \ion{O}{7} emission originating in 2--$3\times10^{6}~{\rm K}$ gas in the Galactic halo.  On the other hand, the observed ratio of \ion{C}{6} Ly$\gamma$ to Ly$\alpha$ is $0.3\pm0.2$.  The expected ratios are 0.04 for thermal emission and 0.24 for charge exchange, indicating that charge exchange must contribute strongly to this line and therefore potentially to the rest of the ROSAT R12 band usually associated with $10^6~{\rm K}$ emission from the Local Hot Bubble.  The limited statistics of this experiment and systematic uncertainties due to the contamination require only $>32\%$ thermal emission for \ion{O}{7} and $>20\%$ from charge exchange for \ion{C}{6} at the 90\% confidence level.  An experimental gold coating on the silicon substrate of the array greatly reduced extraneous signals induced on nearby pixels from cosmic rays passing through the substrate, reducing the triggered event rate by a factor of 15 from a previous flight of the instrument.  
\end{abstract}


\section{Introduction}

The diffuse X-ray background below 1~keV is largely dominated by Galactic emission and is the source of most of our knowledge of the hot phases of the interstellar medium.  Over the past decade, it has been realized that charge exchange between highly ionized metals in the solar wind and interstellar hydrogen and helium passing through the Solar System should be producing X$\mbox{-}$ray emission that can mimic thermal emission from 1--$3\times10^{6}~{\rm K}$ gas \citep{1996Sci...274..205L,1997GeoRL..24..105C,1998LNP...506..121C}.  Models of this solar wind charge exchange (SWCX) emission are limited by lack of detailed knowledge of the atomic physics and the solar wind composition \citep[][and references therein]{2010SSRv..157...57D}, but the best calculations so far indicate SWCX could be providing almost all of the emission observed in the ROSAT R12 band ($\sim120$--284 eV) at low Galactic latitudes and a substantial fraction of the larger intensity seen in this band at high latitudes \citep{2004A&A...418..143L,2009ApJ...696.1517K}.

The R12 band emission was originally thought to come mostly from a ``Local Hot Bubble" (LHB), an irregular volume of $1.0\times10^{6}~{\rm K}$ gas of $\sim100~{\rm pc}$ extent surrounding the Sun.  The observed intensity is roughly uniform at low latitudes with extended parts of the bubble providing large patches that are 2--3 times brighter at higher latitudes.  Additional bright areas come from large patches of $1.0\times10^{6}~{\rm K}$ gas in the Galactic halo and are partially absorbed by the intervening interstellar material \citep[][and references therein]{2000ApJ...543..195K,1998ApJ...493..715S}.  SWCX cannot reproduce the brighter patches, but replacement of all the LHB emission at low latitudes and a similar approximately uniform intensity at high latitudes would drastically change (and complicate) our picture of the local interstellar medium.

While current best estimates of SWCX with interstellar neutrals indicate that it can easily provide essentially all of the approximately uniform R12 band intensity seen at low latitudes, there is considerable astrophysical evidence suggesting that its contribution should be much less.  There have been four all$\mbox{-}$sky surveys in bands approximating ROSAT R12.  These surveys observed the same regions of the sky from different vantage points around the Earth's orbit and at various phases of the solar cycle, yet they all agree on intensity to a few percent \citep{1995ApJ...454..643S}.  Individual observations often see intensity variations, but these are attributed to charge exchange with neutrals in the magnetosheath, a process whose time variability has largely allowed it to be removed from maps of the R12 band intensity \citep{2008A&A...478..575K}.  The expected SWCX variations for interstellar neutrals have yet to be modeled, but the spatial distribution of the solar wind and its heavy ion content are strongly modulated by the solar cycle and the interplanetary emission differs considerably between the ``upstream" and ``downstream" sides of the flow of interstellar neutrals through the Solar System.  

In the ROSAT R45 band ($\sim480$--1200~eV), most of the Galactic emission is from 1.5--$3\times10^{6}~{\rm K}$ gas in the disk and halo with significant contributions from stellar emission in the plane and toward the Galactic center.  Shadowing observations of nearby molecular clouds show that most of this emission comes from distances $>200~{\rm pc}$ \citep{1993ApJ...409L..21S,1997ApJ...484..245K} and therefore SWCX cannot be a major contributor.  However, some individual emission lines, particularly \ion{O}{7} and \ion{O}{8} K lines, provide interesting and useful diagnostics of the poorly$\mbox{-}$understood distribution of this hotter interstellar gas \citep{2009PASJ...61..805Y} and there is both expectation and evidence that SWCX can contribute significantly to these lines \citep{2009ApJ...707..644G}.

Time variations in the solar wind would seemingly allow easy isolation of a SWCX contribution to the diffuse X$\mbox{-}$ray background, but variations in the solar wind flux with solar cycle are partially cancelled by a correlated variation in depletion of \ion{H}{1} exchange targets by photoionization.  One might also expect variation with Solar System viewing geometry.  This has in fact been observed \citep{2009ApJ...697.1214K} and further observations are planned \citep{2011ExA....32...83G}.  However, these variations are a small fraction of the total predicted SWCX emission, so large and uncertain correction factors must be applied to the measured values.

Another avenue for distinguishing SWCX contributions lies in high spectral resolution observations.  The differences are subtle since thermal emission in this temperature range is almost entirely in collisionally$\mbox{-}$excited lines of the same ions that emit X$\mbox{-}$rays after charge exchange in the solar wind.  However, the excitation mechanism is very different.  In thermal emission near collisional ionization equilibrium, even the first excited state is usually several $kT$ above the ground state so there is very little collisional excitation of higher $n$ states.  On the other hand, in charge exchange the electron is initially transferred to a level at approximately the same binding energy as its ground state level in neutral hydrogen or helium.  For the highly ionized metals producing SWCX, this is a high $n$ state and the ensuing radiative cascade can produce relatively large amounts of the higher transitions in the series.  These diagnostics are robust and depend only on atomic physics for given interstellar temperatures and solar wind velocities.  Both observation and atomic theory are difficult in the R12 band where the line spectrum is crowded and most of the emission is from L lines of intermediate metal ions with many electrons.  However, an initial attempt can be made by picking particularly favorable lines at somewhat higher energies where the spectrum is less crowded.

The X$\mbox{-}$ray Quantum Calorimeter (XQC) is a high resolution spectrometer designed to study the diffuse X$\mbox{-}$ray background emission below 1~keV.  Its high spectral resolution is advantageous for investigating spectral differences between SWCX and thermal emission from hot gas, and the \ion{O}{7} and \ion{C}{6} lines have sufficient intensity to be measured in a brief rocket flight.  In this paper we present the results of a sounding rocket flight of XQC.  A discussion of the sounding rocket instrument is in \S~2 and the flight is described in \S~3.  The data reduction is detailed in \S~4 with analysis of science results in \S~5.

\section{The X$\mbox{-}$ray Quantum Calorimeter Sounding Rocket Instrument}

The XQC instrument has been described previously by \citet{2002ApJ...576..188M}.  It employs microcalorimeter detectors cooled to 50~mK by a combination of pumped liquid helium and an adiabatic demagnetization refrigerator.  The energy resolution was 11~eV~FWHM below 1~keV and a mechanical stop defined the $\sim1~{\rm sr}$ field of view.  Here we give a brief overview of significant changes since the previous flight.

\subsection{UV/Vis/IR Blocking Filters}

\begin{figure}  
\plottwo{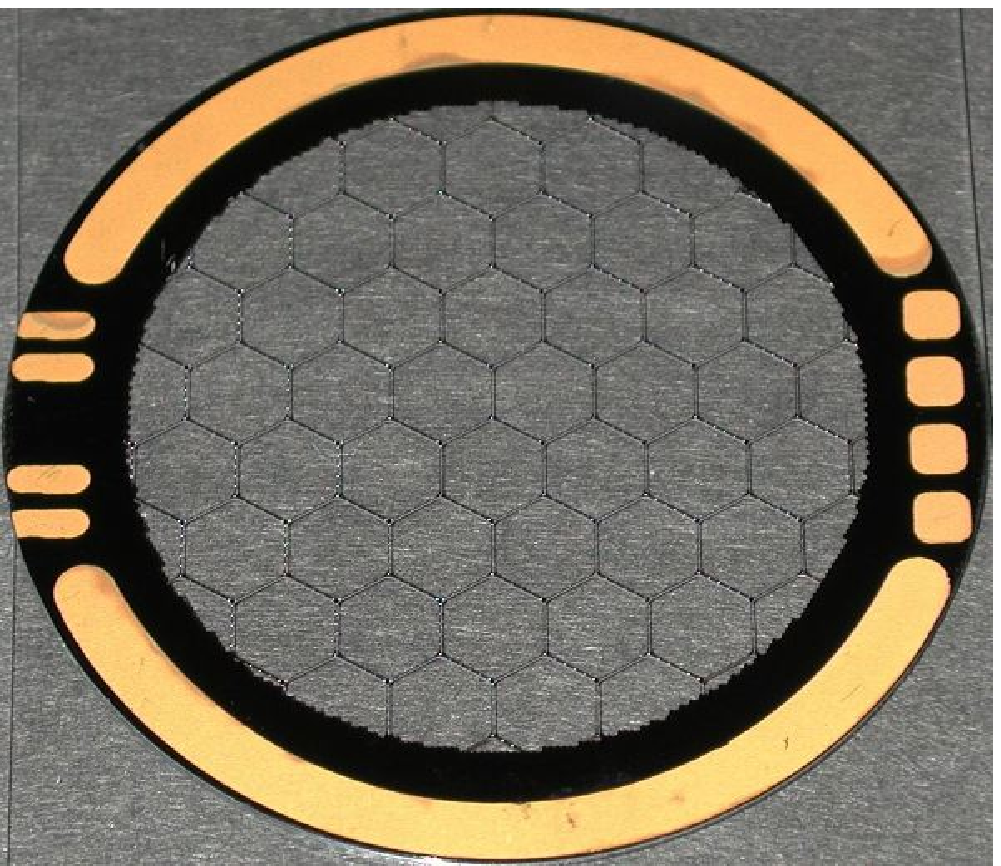}{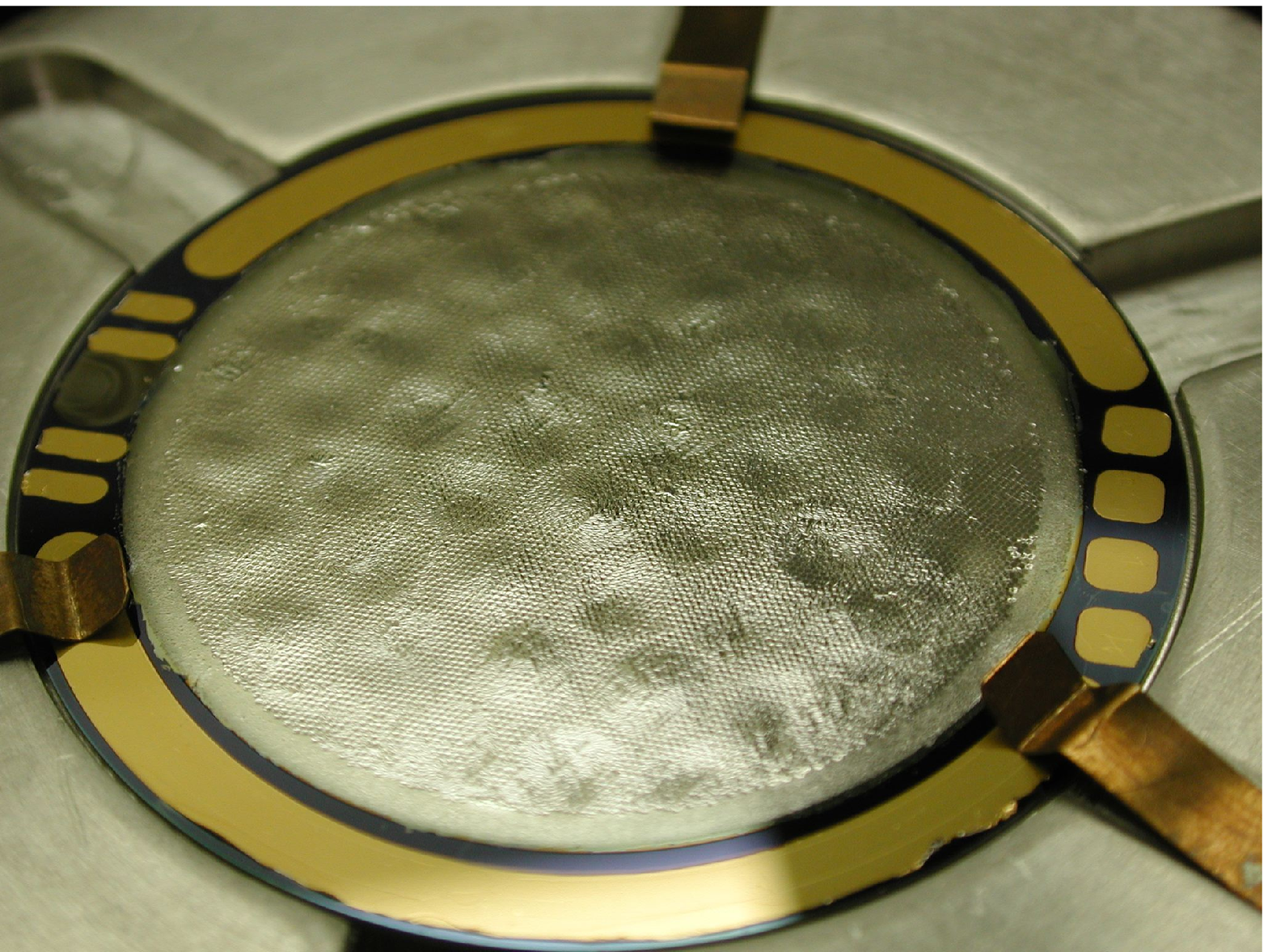}
\caption{Silicon support mesh for the largest of five blocking filters (left).  A 20~nm aluminum filter supported by 50~nm polyimide is bonded to the support mesh (right).  The mesh consists of two layers of hexagonal grids: an 8~$\mu$m thick fine mesh and a 200~$\mu$m thick coarse support mesh.  The fine mesh has 5~$\mu$m leg width and 330~$\mu$m pitch while the coarse mesh has 65~$\mu$m leg width and 5.3~mm pitch.  Only the coarse mesh is visible in the photos.  The transmission of the complete mesh is $\sim97\%$ for X$\mbox{-}$ray energies below 2~keV.\label{fig1}}
\end{figure}

Room$\mbox{-}$temperature infrared radiation must be attenuated by about $10^9$ to limit heat load and shot noise on the detectors.  This was accomplished by a series of five 20~nm aluminum filters supported on thin plastic films.  Two major improvements were made for this flight.  One was to change the film from 100~nm parylene to 50~nm polyimide, improving both strength and X$\mbox{-}$ray transmission.  The other was the addition of support meshes for the three largest filters, a change that increased filter robustness and ease of handling.  The high thermal conductivity of the mesh allowed for better thermal control and more efficient use of decontamination heaters.  The outer mesh and filter are shown in Figure~\ref{fig1}.

\subsection{Calorimeter Array}

\begin{figure} 
\epsscale{.70}
\plotone{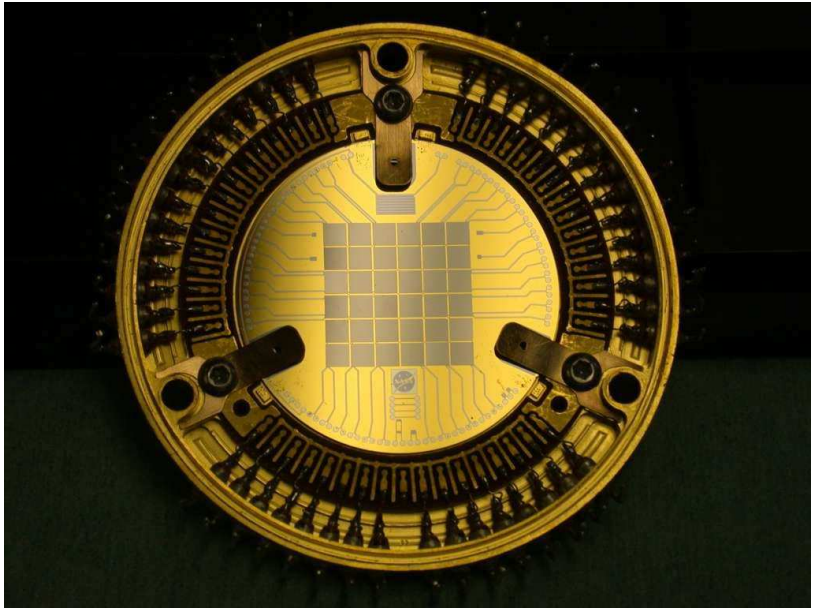}
\caption{Detector array used during flight.  The array, consisting of thirty$\mbox{-}$six $2~{\rm mm}\times2~{\rm mm}$ pixels, is shown mounted in the inner detector box.  The frame was gold$\mbox{-}$coated to minimize response to cosmic ray hits.\label{fig2}}
\end{figure}

The detector array used the same technology employed in the previous flight \citep{2002ApJ...576..188M}: ion$\mbox{-}$implanted semiconductor thermistors with HgTe absorbers.  The new array had thirty$\mbox{-}$six $2~{\rm mm}\times2~{\rm mm}$ pixels in a $6\times6$ configuration, providing four times the collecting area of the previous unit.  The detector time constants were close to 9~ms, about three times longer than intended, which introduced the pileup problems described in \S\ref{opm}.  A 1~$\mu$m gold coating was added to the front of the array's silicon substrate to increase the substrate heat capacity.  This greatly lowered the rate of triggered signals induced on nearby pixels when a cosmic ray passed through the substrate, changing it from $0.1~{\rm counts~s}^{-1}~{\rm mm}^{-2}$ for the previous flight to $0.007~{\rm counts~s}^{-1}~{\rm mm}^{-2}$ for this flight.  It would also greatly alleviate the increase in noise from X$\mbox{-}$ray hits on adjacent pixels and the surrounding frame when a strong source is observed, such as is planned for future X-ray satellite missions.  A photograph of the array is shown in Figure~\ref{fig2}.

\subsection{Electronics}

A few changes were made to the electronics to accommodate the slower pulses.  The sampling rate was halved to 10.4~kHz and the number of samples telemetered on each event was doubled to 2048.  On-board flash memory storage was added to capture the continuous data stream from each pixel, allowing the complete history of all pulses to be taken into account during analysis.

\clearpage
\section{The Sounding Rocket Flight}

The experiment flew 2008~May~1 at 05:30~UT as flight~36.223UG from White Sands Missile Range.  We obtained four minutes of on$\mbox{-}$target data at altitudes above 160~km.  For about 30~minutes immediately before launch and for six minutes afterwards while the payload was descending on the parachute, we recorded calibration data from a low energy fluorescent line source on the gate valve slide.  We centered the field of view at the Galactic coordinates of $l=90\degr{}$, $b=+60\degr{}$, the same target as the previous flight.  A hard landing completely destroyed the detectors and filters, precluding any additional post$\mbox{-}$flight calibration or diagnosis.

\section{Data Reduction}

The data analysis scheme employed an optimal filter, gain drift corrections using pulses from the continuous 3.3~keV calibration source, and individual corrections for the slight energy non$\mbox{-}$linearity of each pixel before combining X$\mbox{-}$ray events into a single spectrum.  The long time constant of these detectors meant that a substantial fraction of the pulses occurred during the tail of a previous pulse, and new data analysis techniques were developed to handle these overlaps.

\subsection{Interference Removal}

Microphonics affected the high$\mbox{-}$impedance detector pixels.  The cryostat was well$\mbox{-}$isolated from the rocket body, but operation of the gate valve motor at the beginning of observations excited strong microphonics due to vibrations of the detector pixels on their thermal isolation support legs.  The vibrations decayed with time constants of seconds to minutes and had fundamental resonant frequencies of around 500~Hz.  The time$\mbox{-}$varying amplitudes of these vibrations were unique to the on$\mbox{-}$target time, so an optimal filter constructed from pre$\mbox{-}$flight data would not have the gate valve induced microphonics and hence would not notch them out.  However, since 98\% of the signal power was below 100~Hz, applying a low pass filter with a corner frequency of $\sim400~{\rm Hz}$ to the data removed the microphonics while having little effect on the signal.

Pickup at 60.04~Hz from a commutator in the housekeeping electronics was well within the signal bandpass, but had a very stable frequency and amplitude.  The interference was fit in the time domain for each pixel and subtracted from the data streams.  This eliminated the interference while removing a negligible amount of signal bandwidth.

\subsection{Overlapped Pulse Technique}\label{opm}

The optimal estimate of the signal amplitude when the shape of the pulse and the noise spectrum are known is obtained with a filter which weights the $i^{\rm th}$ frequency bin by $\hat{s}_i/n_i^2$, where $\hat{s}_i$ is the complex conjugate of the signal power spectrum and $n_i^2$ is the mean noise power \citep{1988ITNS...35...59M,1993JLTP...93..281S,1999SPIE.3765..741B,2005physics...3045M}.  To get sufficient frequency resolution to efficiently take into account $1/f$ noise and interference lines, it is desirable to include about 60~decay time constants in each filtered pulse.

In the original flight electronics, analog filtering and level discriminators were used to trigger the telemetered transmission of a $\sim200~{\rm ms}$ digitized interval around each detected pulse.  These intervals could be digitally filtered on the ground with an optimized filter to extract the best possible signal to noise ratio.  

\begin{figure} 
\plotone{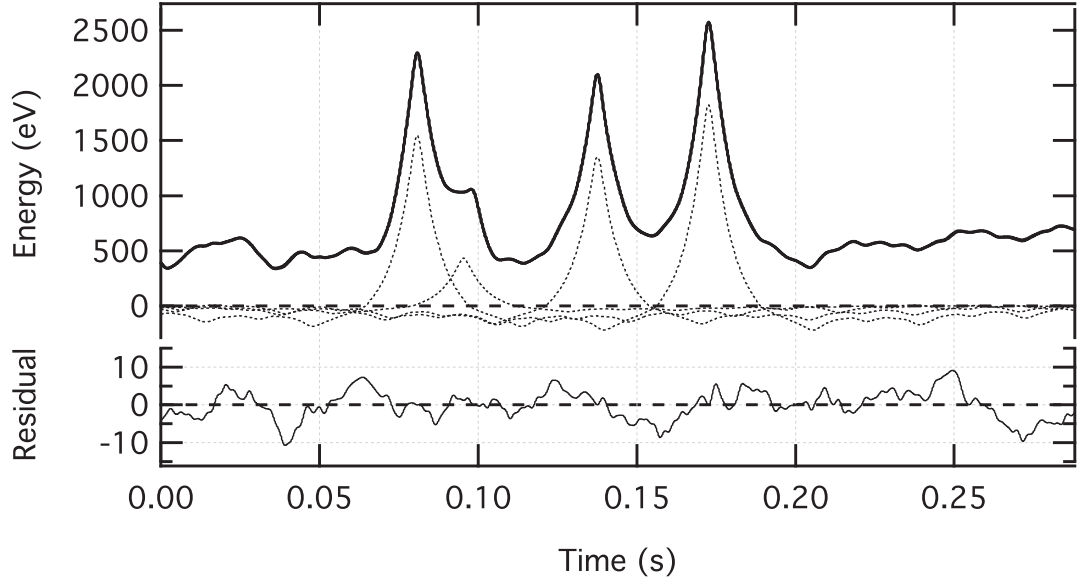}
\caption{Filtered data stream (thick line, offset by 750~eV for clarity) containing four overlapped pulses (top panel).  Models for each of the pulses before summing are plotted with dots.  The residual, also in eV, is the difference between the sum of these and the filtered data stream (bottom panel).  The `wiggles' in the baseline of the filtered pulses are from deterministic ringing of the filter.  The noise level can be seen in the residual and is much smaller.
\label{fig3}}
\end{figure}

Due to slow detectors with time constants of $\sim9~{\rm ms}$ paired with an overall count rate of $1.8~{\rm s}^{-1}~{\rm pixel}^{-1}$ during the on$\mbox{-}$target portion of the flight, there was a high incidence of overlap where the peaks of pulses were separated by less than 600~ms.  To avoid either degraded energy resolution or an unacceptable dead time, we used the continuously logged output of each pixel from the on$\mbox{-}$board data recorder to correct for overlapping pulses.  An optimal filter with 1.3~Hz frequency resolution was constructed by finding enough isolated 3.3~keV calibration line events to make an 800~ms average pulse that was transformed to find the signal power spectrum.  Similarly, we selected pulse$\mbox{-}$free sections to obtain the noise spectrum.  The resulting filter was transformed to the time domain and convolved with the entire data record for that pixel.  We used the filtered average calibration$\mbox{-}$line pulse as a template for the filtered pulse shape.  In the assumption of strict linearity, the filtered record should be the sum of suitably scaled copies of this template at the location of each pulse.  We located the largest pulses first and used an iterative subtraction to locate all the events above noise.  Then a simultaneous fit was made to refine the times and amplitudes of all the located pulses.  An example of fitting four overlapped pulses is given in Figure~\ref{fig3}.  The detectors are actually about 2.5\% nonlinear at 3.3~keV, primarily due to the large nonlinearity of the temperature dependence of the thermometer resistance.  We developed an empirical correction for pulses peaking within 120~ms of a large ($>2.5~{\rm keV}$) pulse.  Shape changes were very small and were neglected.

The on$\mbox{-}$target data were processed with both this method and a conventional one$\mbox{-}$pulse$\mbox{-}$at$\mbox{-}$a$\mbox{-}$time analysis that used a 200~ms sample length to keep dead time low.  The overlapped analysis had an energy resolution of 11~eV~FWHM below 1~keV while the conventional analysis gave 22~eV~FWHM.  For pulses separated by $\le24~{\rm ms}$ where the initial pulse had an energy below 2.5~keV and by $\le72~{\rm ms}$ for initial pulses above 2.5~keV, energy resolution was degraded even after correction and these time intervals were excluded.  Though effective for our limited amount of data, this technique is not an ideal solution as it is computationally expensive and is impractical to run in real time.

\subsection{Blocking Filter Contamination}

By comparing to a previous observation of the same field \citep{2002ApJ...576..188M}, we found that our observed sky rates were roughly 75\% too low at energies below 0.5~keV, $\sim85\%$ too low around 0.6~keV, and $\sim30\%$ too low around 1.5~keV.  Since measurements taken over many years have found similar emission for times not enhanced by solar flares or coronal mass ejections \citep{1990ARA&A..28..657M,1995ApJ...454..643S,2009PASJ...61..805Y}, the lower rates indicated a loss of sensitivity.  This was almost certainly due to contamination on one or more of the infrared blocking filters.

Calibration data confirmed that contamination formed during the flight.  During times preceding the launch and on the parachute, a small fluorescent source on the gate valve slide illuminated the array with low energy lines at energies from 0.15 to 2.1~keV.  These lines were used to check gain linearity and monitor filter transmission before and at the end of the flight.

Two of the gate valve source lines, O~K$\alpha$ and F~K$\alpha$ at 525 and 677~eV respectively, fall on opposite sides of the 543~eV oxygen absorption edge making their ratio a highly sensitive probe of water ice contamination.  The rates were as expected for 387~s of good pre$\mbox{-}$launch data at $\sim0.21$ and $0.25~{\rm counts}~{\rm s}^{-1}~{\rm pixel}^{-1}$.  For 44~s of good data immediately following the observation while the payload was descending on the parachute, the rates had changed to 0.15 and $0.06~{\rm counts}~{\rm s}^{-1}~{\rm pixel}^{-1}$.  This change is consistent with the formation of about 1.1~microns of ice and appears to have occurred shortly after the gate valve was opened at 160~km, since the observed sky flux is constant for the remainder of the observation.  A possible source of the ice was payload outgassing.  This problem had not been observed on previous flights, but modifications for this flight included removal of a hermetic bulkhead separating the experiment sections from the rest of the payload (telemetry, boost$\mbox{-}$phase guidance system, attitude control system, and parachute) so the entire payload was vented through the aft end past the open gate valve.

Detected X$\mbox{-}$rays from the gate valve source pass through only the central 1\% of the area of the outermost filter.  Although ice contamination in the center of the outermost filter was consistent with the changes between pre$\mbox{-}$launch and parachute calibration data, the on$\mbox{-}$target sky rates were down by about a factor of four at most energies below 1~keV which is not compatible with a uniform layer of 1.1~$\mu$m ice across the filter.

\begin{figure} 
\begin{center}
\includegraphics[angle=0,scale=.7]{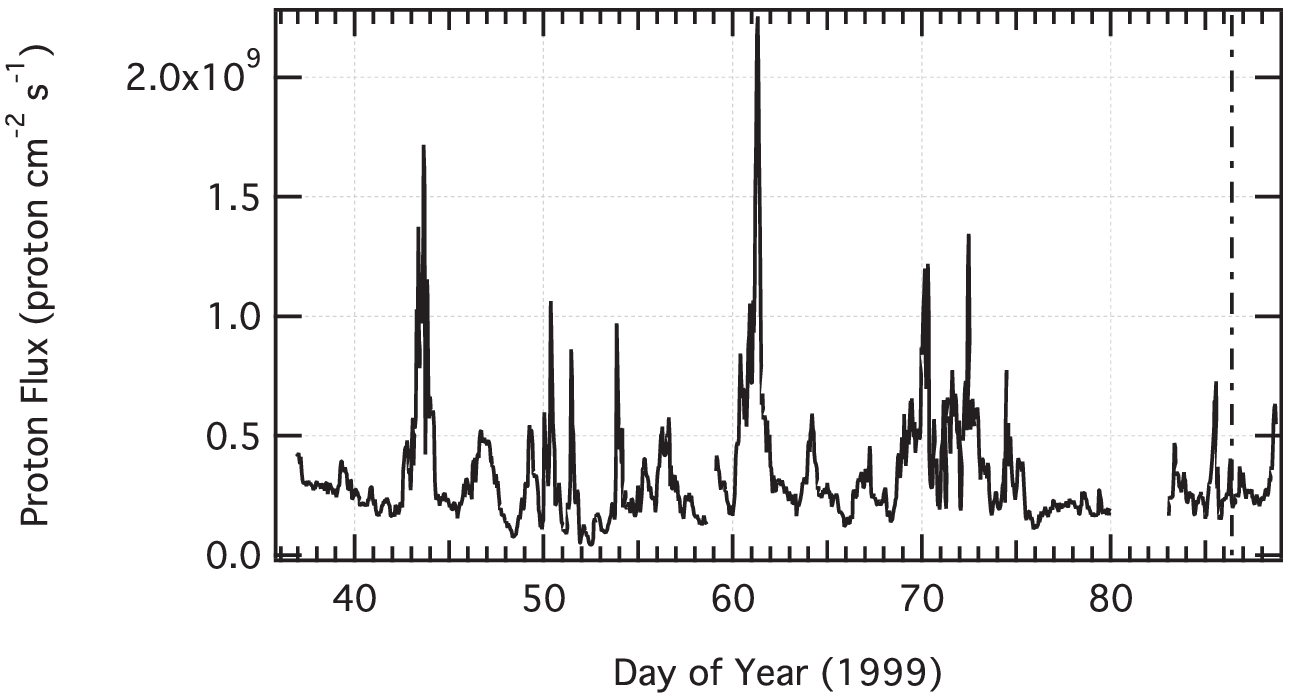} \\
\includegraphics[angle=0,scale=.7]{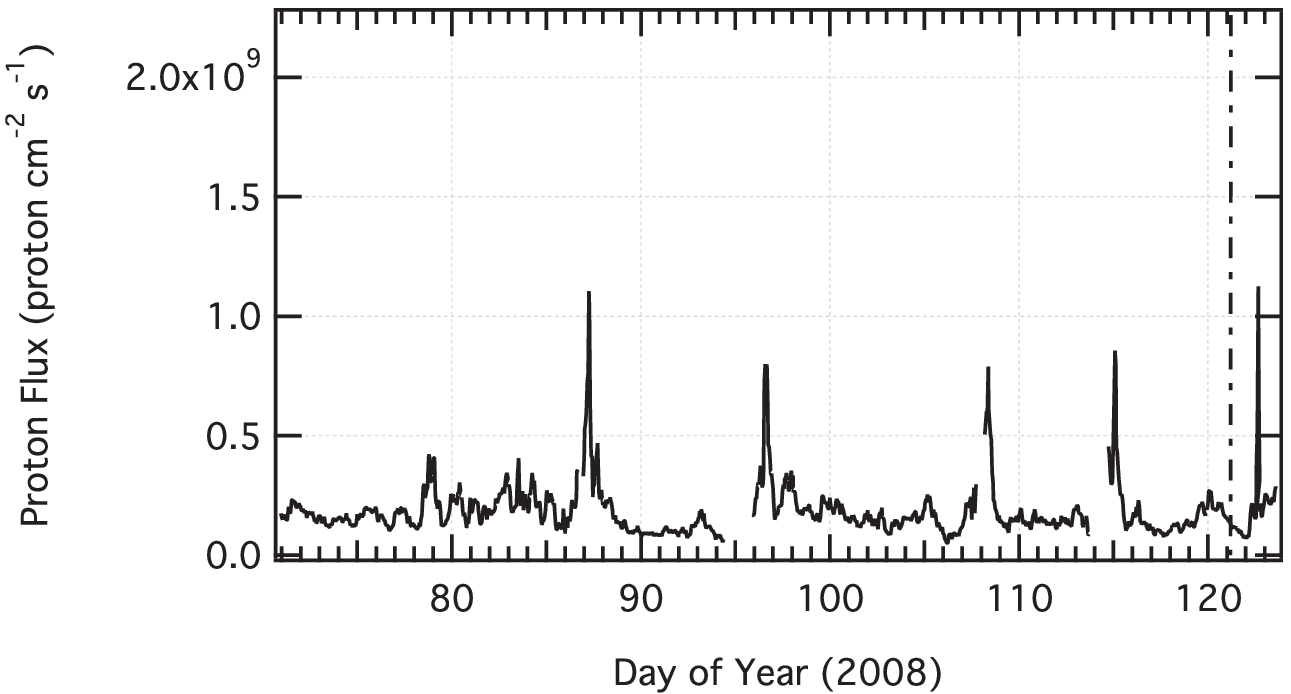}
\end{center}
\caption{Both plots show solar wind proton flux calculated from level~2 SWEPAM data from the ACE satellite and are time shifted to account for propagation time from the satellite to the Earth ($\sim3400~{\rm s}$ for both).  The top graph shows the conditions leading up to flight~27.141 with the launch on day~86.4 (dot$\mbox{-}$dash line).  The bottom graph is for flight~36.223 with the launch on day~121.2 (dot$\mbox{-}$dash line).\label{fig4}}
\end{figure}

To characterize the nature of the contamination over the entire observed solid angle, we compared the on$\mbox{-}$target data to the previous flight.  We fit, using XSPEC version 12.7.0e, a multi$\mbox{-}$temperature APEC model to the data from the previous flight.  The parameters of this model are given in Table~\ref{tbl-1}.  Both flights were made at a similar time of the year, at a similar time in the solar cycle, and with somewhat similar solar wind proton fluxes\footnote{Solar wind data from \newline http://www.srl.caltech.edu/ACE/ASC/level2/index.html} as shown in Figure~\ref{fig4}.

We used the astrophysical source model to predict the observed rates for this flight assuming various models for the contamination.  Since uniform coverage of 1.1~$\mu$m of ice over$\mbox{-}$predicted intensities by a factor of two below the oxygen edge and under$\mbox{-}$predicted by a factor of $\sim1.5$ above the edge, we modeled the ice distribution as consisting of three thicknesses: 1.1~${\mu}$m as seen in the center of the filter at the end of flight, ice thick enough to block sufficient transmission below the oxygen edge ($>15~{\rm microns}$), and areas clear of contamination.  We found the best estimate to be 16\% 1.1~micron ice, 16\% clear of contamination, and 67\% blocked (``model~1").  While this yielded a model that reasonably produced intensities below 0.8~keV, it under$\mbox{-}$predicted by roughly a factor of two above that energy.  Attempts to improve model~1 above 0.8~keV required thinning the blocked section.  This, however, caused over$\mbox{-}$predictions below the oxygen edge.  

Since using an ice$\mbox{-}$only contaminant model could not reproduce the observed spectrum over the entire 0.1--2.0~keV region, we then explored other reasonable contaminants that might have the required behavior.  When we allowed organic (carbon) contamination to replace blocking in the model, the best estimate of contamination coverage was 70\% 880~$\mu{\rm g~cm}^{-2}$ carbon, 20\% 1.1~$\mu$m water ice, and 10\% clear (``model~2").  This produced a reasonable fit over the entire energy range.  A credible fit could also be produced from 81\% 880~$\mu{\rm g~cm}^{-2}$ carbon, 1\% 1.1~$\mu$m water ice, and 18\% clear (``model~3").

Lack of detailed knowledge of the contamination results in large uncertainties in the measured line intensities (roughly a factor of two below 600~eV).  However, due to the small energy difference between \ion{C}{6} Ly$\alpha$ and Ly$\gamma$ and between \ion{O}{7} $F$, $I$ and $R$ lines, there is little effect on these critical line ratios.

\subsection{Derived Pulse Height Spectrum}

\begin{figure} 
\begin{center}
\epsscale{1}
\plotone{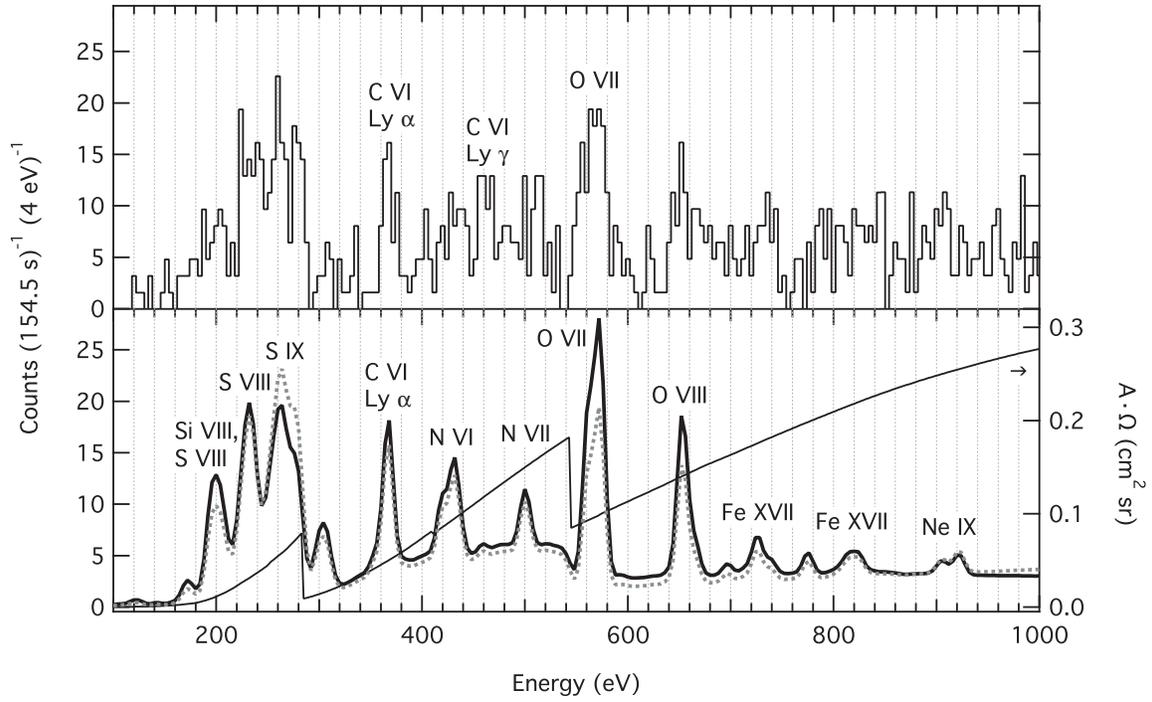}
\end{center}
\caption{Upper panel displays the observed spectrum from the target region of the sky while the lower panel shows the model reported in Table~\ref{tbl-1} folded through both the instrument response and contamination model~1 consisting of regions of thick ice, 1.1~$\mu$m ice, and clear (thick line).  The throughput curve (thin line) in the bottom panel is shaped predominantly by the blocking filters and the contamination model.  For comparison, the bottom panel also shows the folded model for contamination model~2 (dotted line).\label{fig5}}
\end{figure}

While the telemetered on$\mbox{-}$target data had instrumental dead times of $\sim30\%$, there was no instrumental dead time for data from the on$\mbox{-}$board recorders.  Removing time intervals affected by electronically saturating or highly nonlinear events ($\gtrsim5~{\rm keV}$), saturating microphonics, and pileup for close pulses left a net live time of 154.5~s or 60\% of the total.  The spectrum from the target region of the sky is shown in Figure~\ref{fig5} along with the previously described multi$\mbox{-}$temperature thermal emission plus AGN spectrum fit to a previous observation of this region (as given in Table~\ref{tbl-1}) folded through contamination model~1.  The \ion{C}{6} Ly$\alpha$ and Ly$\gamma$ lines are clearly present in the data along with the \ion{O}{8} Ly$\alpha$ line and the helium$\mbox{-}$like \ion{O}{7} triplet.  No useful limits can be placed on the Fe$\mbox{-}$M line complex at 70~eV: only one count is observed, but even very small amounts of uniform ice contamination would greatly reduce sensitivity at such low energies.  

\section{Analysis of Charge Exchange Versus Thermal Emission}
 
\begin{figure} 
\begin{center}
\epsscale{.7}
\plotone{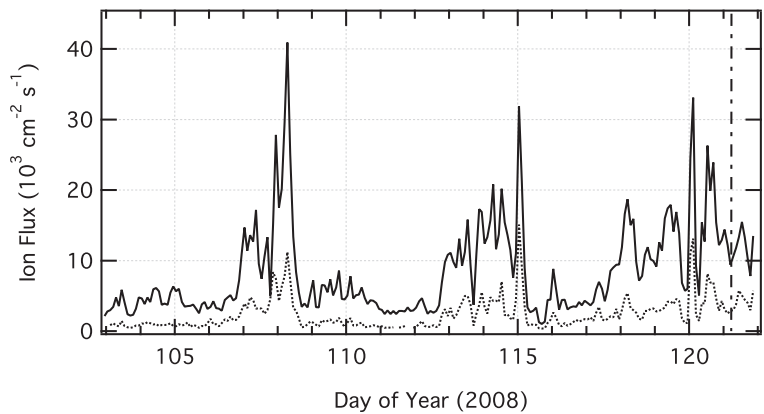}
\end{center}
\caption{Heavy ion fluxes for ${\rm C}^{+6}$ (solid line) and ${\rm O}^{+7}$ (dotted line) are calculated using data from the ACE SWICS instrument and are shifted in time by $\sim3400~{\rm s}$ to correct for travel time from the satellite to the Earth.  The dot$\mbox{-}$dash line shows the time of flight.\label{fig6}}
\end{figure}

\begin{figure} 
\begin{center}
\includegraphics[angle=0,scale=0.161]{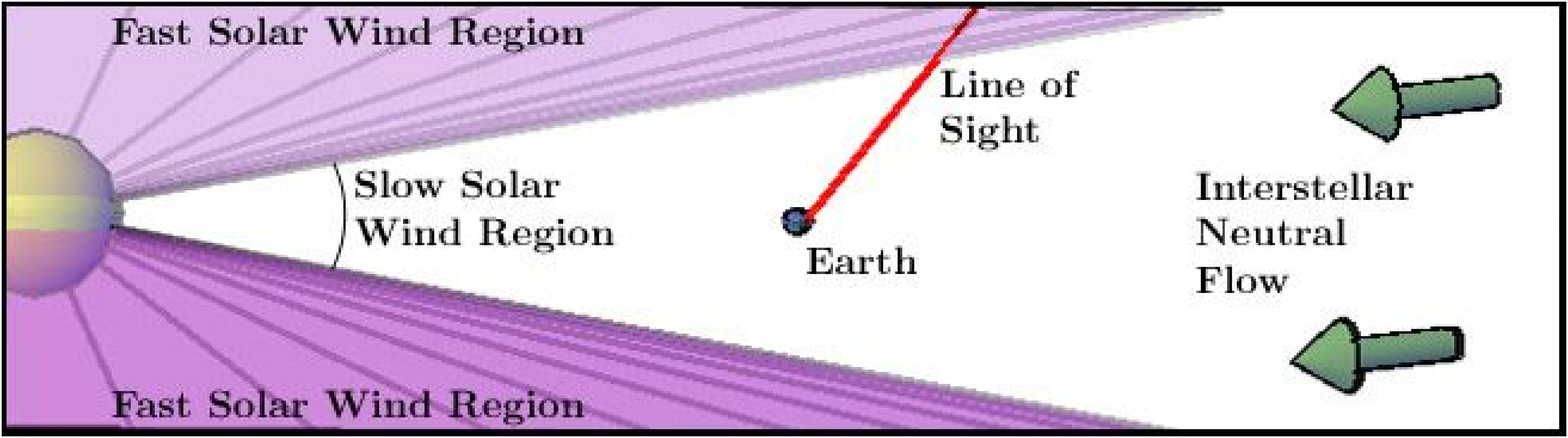} \\
\includegraphics[angle=0,scale=0.35]{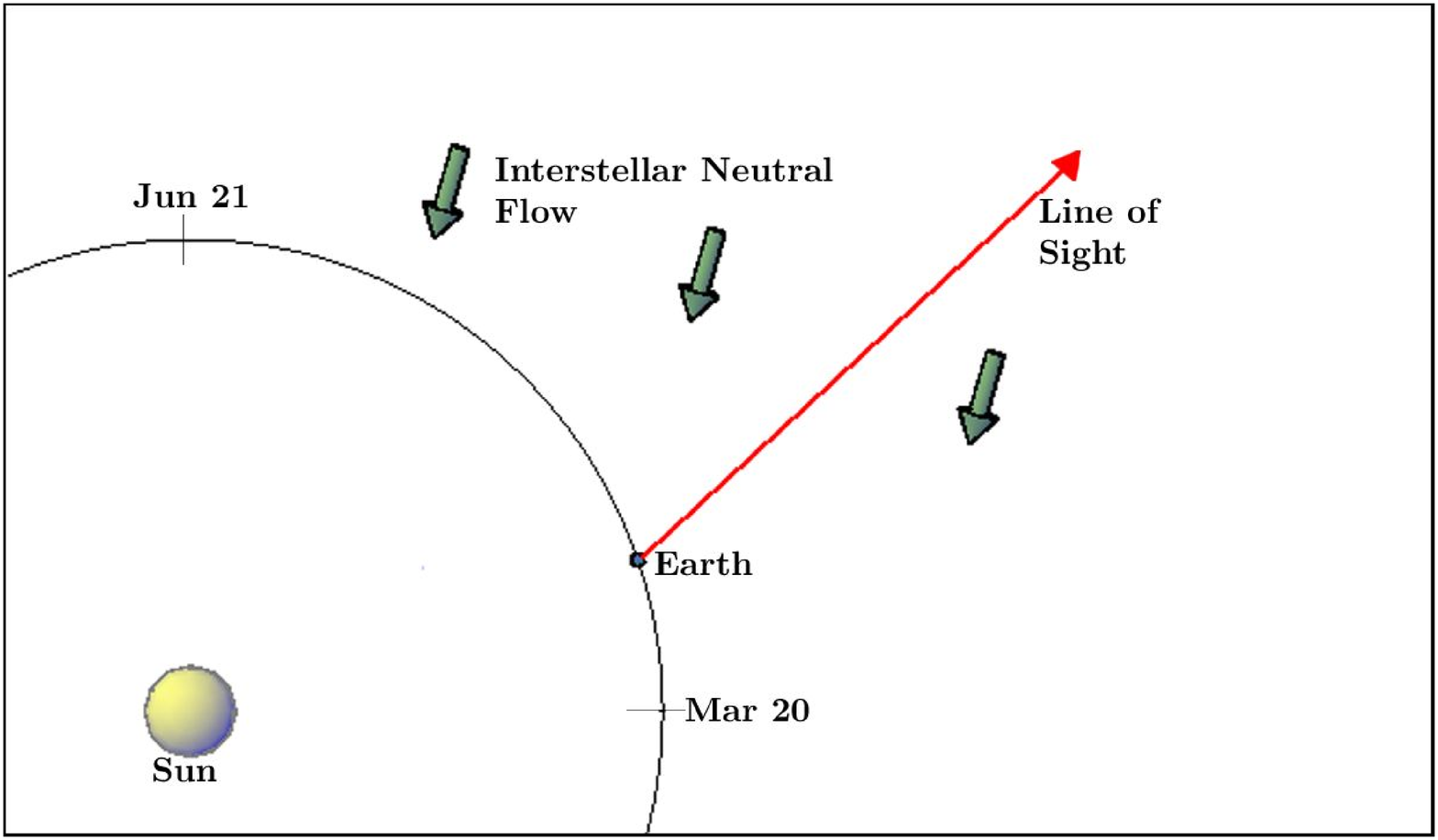}
\caption{Line of sight for the observation from a side view (upper panel) and a top view (lower panel).\label{fig7}}
\end{center}
\end{figure}

The proton flux shown in Figure~\ref{fig4} and the heavy ion fluxes in Figure~\ref{fig6} were at values consistent with times not enhanced by solar flares or coronal mass ejections for the five days preceding flight \citep[][and references therein]{2010ApJS..187..388H,2006A&A...460..289K}.  In these circumstances, charge exchange on neutrals in the Earth's geocorona should be negligible \citep{2003JGRA..108.8031R}.  Our high ecliptic latitude line of sight had a short path through the slow solar wind, which has the larger heavy ion content \citep{2009SSRv..143..217K}, and passed mostly through regions of fast wind as seen in Figure~\ref{fig7}.  For this sight$\mbox{-}$line with average solar minimum conditions, D. Koutroumpa (2012, private communication) used the method outlined in \citet{2006A&A...460..289K} to predict \ion{O}{7} intensities of $F=0.80~{\rm photons}~{\rm cm}^{-2}~{\rm s}^{-1}~{\rm sr}^{-1}$ (LU), $I=0.20~{\rm LU}$, and $R=0.16~{\rm LU}$, giving a mean energy of 564.2~eV.  She also predicted \ion{C}{6} intensities of ${\rm Ly}\alpha=1.94~{\rm LU}$ and ${\rm Ly}\gamma=0.47~{\rm LU}$ for a ratio of ${\rm Ly}\gamma/{\rm Ly}\alpha=0.24$.  The fractional line contributions for thermal emission were obtained from the APEC thermal models in XSPEC as shown in Table~\ref{tbl-2}.  As can be seen from the table, the derived line contributions are not sensitive to details of the thermal models.

Line fluxes were initially determined by setting the C and O abundances to zero in the XSPEC thermal model of Table~\ref{tbl-1}, but otherwise keeping all parameters fixed.  Emission for C and O was supplied by lines of fixed energy and negligible width but adjustable amplitude for \ion{C}{6} Ly$\alpha$ (368~eV), \ion{C}{6} Ly$\gamma$ (459~eV), \ion{O}{8} (653~eV), and \ion{O}{7} K$\beta$ (666~eV).  We did not attempt to fit \ion{C}{6} Ly$\beta$ because it is blended with stronger \ion{N}{6} lines.  The \ion{O}{7} triplet ($F=560.9~{\rm eV}$, $I=568.5~{\rm eV}$, and $R=574.0~{\rm eV}$) was fit with a single Gaussian of variable energy, amplitude, and resolution.  Table~\ref{tbl-3} gives the \ion{C}{6} line ratios and \ion{O}{7} K$\mbox{-}$line centroid (mean of Gaussian) energies for the different contamination models, along with the values expected for pure thermal and charge exchange sources.  The observed values are consistent with \ion{C}{6} being dominated by charge exchange, while \ion{O}{7} appears to be predominantly thermal emission.  The relatively large uncertainties require only that charge exchange provide at least 0.2, 0.3, and 0.6 of the total \ion{C}{6} emission for the different contamination models, while it generates less than 0.81, 0.76, and 0.88 of the \ion{O}{7}.  Confidence limits were derived by varying the \ion{C}{6} line ratio or \ion{O}{7} Gaussian mean energy until $\chi^2$ was increased by 2.706 above its minimum value (90\% confidence for a single parameter), while simultaneously minimizing with respect to all other parameters.  We found energy scale systematic uncertainties to be small: fluorescent lines at 277, 525, and 677~eV from the gate valve slide calibration source were all within 0.5~eV of the expected energies during the 30~minutes directly preceding the launch, and the 3313~eV continuous source used to track the detector gain showed that it changed by less than the $\sim0.2\%$ statistical uncertainty between this period and the on$\mbox{-}$target time.

The line ratio and centroid values shown in Table~\ref{tbl-3} are themselves largely independent of any assumptions about the source spectrum and could be used to constrain different models.  However, our energy resolution is comparable to the \ion{O}{7}~K line spacing, so it should be possible to get a better limit on the fractional contribution of charge exchange if we use the predicted line ratios and fit the individual lines.  We again took the model from Table~\ref{tbl-1} with fixed parameters, set the O and C abundances to zero, and substituted fixed$\mbox{-}$energy lines for these elements.  We fit six lines, $F_{\rm CX}$, $I_{\rm CX}$, $R_{\rm CX}$, $F_{\rm THERM}$, $I_{\rm THERM}$, and $R_{\rm THERM}$, with ratios constrained by the predictions shown in Table~\ref{tbl-2}, so the six lines could be fit with two free parameters, taken to be the sum of the six components and the ratio of the sum of the three charge exchange lines to all six.  The ratio of $F_{\rm CX}+I_{\rm CX}+R_{\rm CX}$ to the total thermal plus charge exchange \ion{O}{7} was varied while simultaneously minimizing other parameters until $\chi^2$ increased by 2.706 above its minimum.  The resulting 90\% upper limits on the fraction are 0.68, 0.65, and 0.67 for the three contamination models, as shown in Table~\ref{tbl-4}.  These are a noticeable improvement over the limits derived from the mean energy fits in Table~\ref{tbl-3}.  For consistency we used an equivalent procedure to derive limits on the charge exchange contribution to \ion{C}{6}, although we do not expect to get different results in this case and in fact the limits in Table~\ref{tbl-4} are identical to those derived from the line ratio limits in Table~\ref{tbl-3}.  

We can combine the total \ion{O}{7} line intensity of 4.8~LU from the previous flight \citep{2002ApJ...576..188M} with our fractional SWCX contribution to get 1.1~{\rm LU} for the heliospheric SWCX \ion{O}{7}~K$\alpha$ line during the current observation.  This is consistent with Koutroumpa's \ion{O}{7} prediction of 1.2~LU of heliospheric SWCX for this line of sight during solar minimum.    Foreground ($D<~200~{\rm pc}$) emission from shadowing observations at high ecliptic latitudes during solar minimum is $\sim1\pm1~{\rm LU}$ \citep{2009ApJ...707..644G,2008ApJ...676..335H}.  \citet{YoshPhD} observed \ion{O}{7} and \ion{O}{8} emission at many latitudes during solar minimum and found there to be an approximately isotropic $2\pm1~{\rm LU}$ floor in \ion{O}{7} intensity that had no associated \ion{O}{8}.  He attributed this to either steady heliospheric charge exchange or thermal emission from the LHB.  \citet{2010AAS...21541524S} however put a two sigma upper limit of 1~LU on any \ion{O}{7} emission from the LHB based on lack of spatial correlation between \ion{O}{7} flux and R12 band emission.  These observations provide some evidence for a SWCX contribution that is similar to our findings.
 
\section{Conclusions}

Despite the loss of sensitivity due to contamination of a blocking filter, we can conclude that there is a significant contribution of: 1) thermal emission to the observed \ion{O}{7} emission and 2) SWCX to \ion{C}{6} emission.  Since the significance of these results is limited almost entirely by statistical precision, modest amounts of additional data will appreciably improve our understanding of SWCX contributions to these lines.  Better limits on contributions to the bulk of the flux in the important R12 band where emission lines are narrowly spaced will require both improved instrumentation to resolve individual lines below 280~eV and increased knowledge of the atomic physics of both charge exchange and collisional excitation for some of the important transitions.

\acknowledgments

Many graduate and undergraduate students have been involved with the development and improvement of this instrument and we greatly appreciate their assistance.  We would particularly like to thank Greg Jaehnig for his contributions.  We also thank Kari Kripps, Regis Brekosky, and John Gygax for their work on the fabrication and assembly of the filters and detectors.  We appreciate Dimitra Koutroumpa providing us with charge exchange predictions of \ion{C}{6} and \ion{O}{7} for our line of sight.  We thank the Luxel Corporation for developing the polyimide filters.  We are grateful for the support of the sounding rocket staff at Wallops Flight Facility and White Sands Missile Range.  We thank the anonymous referee for suggestions that have significantly improved the clarity of the paper.  This work was supported in part by NASA grant NNX09AF09G.

\clearpage
\bibliography{article_Crowder}

\begin{thebibliography}{31}
\expandafter\ifx\csname natexlab\endcsname\relax\def\natexlab#1{#1}\fi

\bibitem[{{Anders} \& {Grevesse}(1989)}]{1989GeCoA..53..197A}
{Anders}, E., \& {Grevesse}, N. 1989, \gca, 53, 197

\bibitem[{{Boyce} {et~al.}(1999){Boyce}, {Audley}, {Baker}, {Dumonthier},
  {Fujimoto}, {Gendreau}, {Ishisaki}, {Kelley}, {Stahle}, {Szymkowiak}, \&
  {Winkert}}]{1999SPIE.3765..741B}
{Boyce}, K.~R., {Audley}, M.~D., {Baker}, R.~G., {et~al.} 1999, in SPIE
  Conference Series, Vol. 3765, SPIE Conference Series, ed. {O.~H.~Siegmund \&
  K.~A.~Flanagan} (SPIE), 741--750

\bibitem[{{Cox}(1998)}]{1998LNP...506..121C}
{Cox}, D.~P. 1998, in Lecture Notes in Physics, Berlin Springer Verlag, Vol.
  506, IAU Colloq. 166: The Local Bubble and Beyond, ed. {D.~Breitschwerdt,
  M.~J.~Freyberg, \& J.~Truemper} (Springer Berlin / Heidelberg), 121--131

\bibitem[{{Cravens}(1997)}]{1997GeoRL..24..105C}
{Cravens}, T.~E. 1997, \grl, 24, 105

\bibitem[{{Dennerl}(2010)}]{2010SSRv..157...57D}
{Dennerl}, K. 2010, \ssr, 157, 57

\bibitem[{{Galeazzi} {et~al.}(2011){Galeazzi}, {Chiao}, {Collier}, {Cravens},
  {Koutroumpa}, {Kuntz}, {Lepri}, {McCammon}, {Porter}, {Prasai}, {Robertson},
  {Snowden}, \& {Uprety}}]{2011ExA....32...83G}
{Galeazzi}, M., {Chiao}, M., {Collier}, M.~R., {et~al.} 2011, Experimental
  Astronomy, 32, 83

\bibitem[{{Gupta} {et~al.}(2009){Gupta}, {Galeazzi}, {Koutroumpa}, {Smith}, \&
  {Lallement}}]{2009ApJ...707..644G}
{Gupta}, A., {Galeazzi}, M., {Koutroumpa}, D., {Smith}, R., \& {Lallement}, R.
  2009, \apj, 707, 644

\bibitem[{{Henley} \& {Shelton}(2008)}]{2008ApJ...676..335H}
{Henley}, D.~B., \& {Shelton}, R.~L. 2008, \apj, 676, 335

\bibitem[{{Henley} \& {Shelton}(2010)}]{2010ApJS..187..388H}
---. 2010, \apjs, 187, 388

\bibitem[{{Koutroumpa} {et~al.}(2009{\natexlab{a}}){Koutroumpa}, {Collier},
  {Kuntz}, {Lallement}, \& {Snowden}}]{2009ApJ...697.1214K}
{Koutroumpa}, D., {Collier}, M.~R., {Kuntz}, K.~D., {Lallement}, R., \&
  {Snowden}, S.~L. 2009{\natexlab{a}}, \apj, 697, 1214

\bibitem[{{Koutroumpa} {et~al.}(2009{\natexlab{b}}){Koutroumpa}, {Lallement},
  {Kharchenko}, \& {Dalgarno}}]{2009SSRv..143..217K}
{Koutroumpa}, D., {Lallement}, R., {Kharchenko}, V., \& {Dalgarno}, A.
  2009{\natexlab{b}}, \ssr, 143, 217

\bibitem[{{Koutroumpa} {et~al.}(2006){Koutroumpa}, {Lallement}, {Kharchenko},
  {Dalgarno}, {Pepino}, {Izmodenov}, \& {Qu{\'e}merais}}]{2006A&A...460..289K}
{Koutroumpa}, D., {Lallement}, R., {Kharchenko}, V., {et~al.} 2006, \aap, 460,
  289

\bibitem[{{Koutroumpa} {et~al.}(2009{\natexlab{c}}){Koutroumpa}, {Lallement},
  {Raymond}, \& {Kharchenko}}]{2009ApJ...696.1517K}
{Koutroumpa}, D., {Lallement}, R., {Raymond}, J.~C., \& {Kharchenko}, V.
  2009{\natexlab{c}}, \apj, 696, 1517

\bibitem[{{Kuntz} \& {Snowden}(2000)}]{2000ApJ...543..195K}
{Kuntz}, K.~D., \& {Snowden}, S.~L. 2000, \apj, 543, 195

\bibitem[{{Kuntz} \& {Snowden}(2008)}]{2008A&A...478..575K}
---. 2008, \aap, 478, 575

\bibitem[{{Kuntz} {et~al.}(1997){Kuntz}, {Snowden}, \&
  {Verter}}]{1997ApJ...484..245K}
{Kuntz}, K.~D., {Snowden}, S.~L., \& {Verter}, F. 1997, \apj, 484, 245

\bibitem[{{Lallement}(2004)}]{2004A&A...418..143L}
{Lallement}, R. 2004, \aap, 418, 143

\bibitem[{{Lisse} {et~al.}(1996){Lisse}, {Dennerl}, {Englhauser}, {Harden},
  {Marshall}, {Mumma}, {Petre}, {Pye}, {Ricketts}, {Schmitt}, {Trumper}, \&
  {West}}]{1996Sci...274..205L}
{Lisse}, C.~M., {Dennerl}, K., {Englhauser}, J., {et~al.} 1996, Science, 274,
  205

\bibitem[{{McCammon}(2005)}]{2005physics...3045M}
{McCammon}, D. 2005, in Cryogenic Particle Detection, ed. C.~Enss
  (Springer-Verlag), 1--34

\bibitem[{{McCammon} \& {Sanders}(1990)}]{1990ARA&A..28..657M}
{McCammon}, D., \& {Sanders}, W.~T. 1990, \araa, 28, 657

\bibitem[{{McCammon} {et~al.}(2002){McCammon}, {Almy}, {Apodaca}, {Bergmann
  Tiest}, {Cui}, {Deiker}, {Galeazzi}, {Juda}, {Lesser}, {Mihara},
  {Morgenthaler}, {Sanders}, {Zhang}, {Figueroa-Feliciano}, {Kelley},
  {Moseley}, {Mushotzky}, {Porter}, {Stahle}, \&
  {Szymkowiak}}]{2002ApJ...576..188M}
{McCammon}, D., {Almy}, R., {Apodaca}, E., {et~al.} 2002, \apj, 576, 188

\bibitem[{{Moseley} {et~al.}(1988){Moseley}, {Kelley}, {Schoelkopf},
  {Szymkowiak}, \& {McCammon}}]{1988ITNS...35...59M}
{Moseley}, S.~H., {Kelley}, R.~L., {Schoelkopf}, R.~J., {Szymkowiak}, A.~E., \&
  {McCammon}, D. 1988, IEEE Transactions on Nuclear Science, 35, 59

\bibitem[{{Robertson} \& {Cravens}(2003)}]{2003JGRA..108.8031R}
{Robertson}, I.~P., \& {Cravens}, T.~E. 2003, Journal of Geophysical Research
  (Space Physics), 108, 8031

\bibitem[{{Savage} \& {Sembach}(1996)}]{1996ARA&A..34..279S}
{Savage}, B.~D., \& {Sembach}, K.~R. 1996, \araa, 34, 279

\bibitem[{{Snowden} {et~al.}(1998){Snowden}, {Egger}, {Finkbeiner}, {Freyberg},
  \& {Plucinsky}}]{1998ApJ...493..715S}
{Snowden}, S.~L., {Egger}, R., {Finkbeiner}, D.~P., {Freyberg}, M.~J., \&
  {Plucinsky}, P.~P. 1998, \apj, 493, 715

\bibitem[{{Snowden} {et~al.}(1993){Snowden}, {McCammon}, \&
  {Verter}}]{1993ApJ...409L..21S}
{Snowden}, S.~L., {McCammon}, D., \& {Verter}, F. 1993, \apjl, 409, L21

\bibitem[{{Snowden} {et~al.}(1995){Snowden}, {Freyberg}, {Plucinsky},
  {Schmitt}, {Truemper}, {Voges}, {Edgar}, {McCammon}, \&
  {Sanders}}]{1995ApJ...454..643S}
{Snowden}, S.~L., {Freyberg}, M.~J., {Plucinsky}, P.~P., {et~al.} 1995, \apj,
  454, 643

\bibitem[{{Swaney} {et~al.}(2010){Swaney}, {Fujimoto}, {Hagihara}, {McCammon},
  {Mitsuda}, {Takei}, {Wang}, {Yao}, \& {Yoshino}}]{2010AAS...21541524S}
{Swaney}, M., {Fujimoto}, R., {Hagihara}, T., {et~al.} 2010, in \baas, Vol.~42,
  AAS Meeting Abstracts \#215 (AAS), 415.24

\bibitem[{{Szymkowiak} {et~al.}(1993){Szymkowiak}, {Kelley}, {Moseley}, \&
  {Stahle}}]{1993JLTP...93..281S}
{Szymkowiak}, A.~E., {Kelley}, R.~L., {Moseley}, S.~H., \& {Stahle}, C.~K.
  1993, Journal of Low Temperature Physics, 93, 281

\bibitem[{{Yoshino}(2008)}]{YoshPhD}
{Yoshino}, T. 2008, PhD thesis, University of Tokyo

\bibitem[{{Yoshino} {et~al.}(2009){Yoshino}, {Mitsuda}, {Yamasaki}, {Takei},
  {Hagihara}, {Masui}, {Bauer}, {McCammon}, {Fujimoto}, {Wang}, \&
  {Yao}}]{2009PASJ...61..805Y}
{Yoshino}, T., {Mitsuda}, K., {Yamasaki}, N.~Y., {et~al.} 2009, \pasj, 61, 805

\end{thebibliography}

\clearpage


\clearpage
\begin{deluxetable}{lcccc} 
\tablecaption{Model Parameters for Fit to XQC Flight 27.141\label{tbl-1}} \tablewidth{0pt}
\tablehead{
\colhead{} & \colhead{$kT$} & \colhead{Emission Measure} & \colhead{} & \colhead{\emph{N}$_\mathrm{H}$} \\
\colhead{Component} & \colhead{(keV)} & \colhead{(${\rm cm}^{-6}~{\rm pc}$)} & \colhead{Abundance} & \colhead{($10^{20}~{\rm H~atoms~cm}^{-2}$)}
}
\startdata
Unabsorbed thermal\tablenotemark{a} & 0.0808 & 0.0140 & Depleted\tablenotemark{b} & ... \\
Absorbed thermal\tablenotemark{a} & 0.0808 & 0.0087 & Solar\tablenotemark{c} & 1.8 \\
Absorbed thermal\tablenotemark{a} & 0.188 & 0.0027 & Solar\tablenotemark{c} & 1.8 \\
Absorbed power$\mbox{-}$law\tablenotemark{d} & ... & ... & ... & 1.8 
\enddata
\tablenotetext{a}{The thermal components use the vvapec model in XSPEC.}
\tablenotetext{b}{From \citet{1996ARA&A..34..279S} for cool clouds toward $\zeta$ Ophiuchi.}
\tablenotetext{c}{\citet{1989GeCoA..53..197A}.}
\tablenotetext{d}{To approximate the composite spectra of the faint and bright active galactic nuclei resolved by Chandra but not removed from this observation, the following model was used below 0.9~keV: $5.7E^{-1.54}+4.9E^{-1.96}$.  Above 0.9~keV the model was: $11E^{-1.4}$. (R. Mushotzky 2001, private communication)}
\end{deluxetable}

\clearpage
\begin{deluxetable}{lccccc} 
\setlength{\tabcolsep}{0.02in} 
\tablecaption{Predicted Fractional Components\label{tbl-2}} \tablewidth{0pt}
\tablehead{
\colhead{} & \colhead{\ion{O}{7} } & \colhead{\ion{O}{7} } & \colhead{\ion{O}{7} } & \colhead{\ion{C}{6} } & \colhead{\ion{C}{6} } \\
\colhead{Component} & \colhead{$F/(F+I+R)$} & \colhead{$I/(F+I+R)$} & \colhead{$R/(F+I+R)$} & \colhead{Ly$\alpha/({\rm Ly}\alpha+{\rm Ly}\gamma$)} & \colhead{Ly$\gamma/({\rm Ly}\alpha+{\rm Ly}\gamma$)}
} 
\startdata
Unabsorbed thermal (0.0808~keV)\tablenotemark{a} & 0.42 & 0.10 & 0.48 & 0.96 & 0.04 \\
Absorbed thermal (0.0808~keV)\tablenotemark{a} & 0.42 & 0.10 & 0.48 & 0.97 & 0.03\\
Absorbed thermal (0.188~keV)\tablenotemark{a} & 0.38 & 0.09 & 0.53 & 0.94 & 0.06\\
\hline
Net thermal & 0.39 & 0.09 & 0.51 & 0.96 & 0.04\\
SWCX & 0.69 & 0.17 & 0.14 & 0.80 & 0.20
\enddata
\tablenotetext{a}{Thermal components described in Table~\ref{tbl-1}.}
\end{deluxetable}

\clearpage
\begin{deluxetable}{lccccc} 
\tablecaption{Line Ratios and Energies\label{tbl-3}} \tablewidth{0pt}
\tablehead{
\colhead{} & \colhead{Contaminant Model} & \colhead{Contaminant Model} & \colhead{Contaminant Model} & \colhead{Expected} & \colhead{Expected} \\
\colhead{Identification} & \colhead{1\tablenotemark{a}} & \colhead{2\tablenotemark{b}} & \colhead{3\tablenotemark{c}} & \colhead{SWCX} & \colhead{Thermal}  
}
\startdata
\ion{C}{6}~Ly$\gamma/{\rm Ly}\alpha$ $^{\rm d}$
& 
$0.27\pm0.20$ 
& 
$0.28\pm0.19$ 
&
$0.33\pm0.19$  
&
0.24
&
0.04 \\
\ion{O}{7} K$\alpha$ centroid (eV) $^{\rm e}$
&
$567.9^{+2.4}_{-2.9}$
& 
$568.1^{+2.3}_{-2.9}$
  &
$567.6^{+2.6}_{-2.9}$
&
564.2
&
568.4 \\
\enddata
\tablecomments{All uncertainties and limits are reported at 90\% confidence level.}
\tablenotetext{a}{67\% blocked, 16\% 1.1~$\mu$m water ice, 16\% clear.}
\tablenotetext{b}{70\% 4~$\mu$m carbon, 20\% 1.1~$\mu$m water ice, 10\% clear.}
\tablenotetext{c}{81\% 4~$\mu$m carbon, 1\% 1.1~$\mu$m water ice, 18\%~clear.}
\tablenotetext{d}{Fit to previous flight is $0.2^{+0.3}_{-0.2}$.  \citep{2002ApJ...576..188M}}
\tablenotetext{e}{Fit to previous flight is $568.9^{+5.1}_{-5.3}~{\rm eV}$.}
\end{deluxetable}

\clearpage
\begin{deluxetable}{lccc} 
\tablecaption{Limits on Fractional Contribution of Solar Wind Charge Exchange\label{tbl-4}} \tablewidth{0pt}
\tablehead{
\colhead{} & \colhead{Contaminant Model} & \colhead{Contaminant Model} & \colhead{Contaminant Model} \\
\colhead{Identification} & \colhead{1\tablenotemark{a}} & \colhead{2\tablenotemark{a}} & \colhead{3\tablenotemark{a}} 
}
\startdata
\ion{C}{6}$_{\rm SWCX}/$\ion{C}{6}$_{\rm Total}$  
  &
$>0.2$
& 
$>0.3$
  &
$>0.6$ \\
\ion{O}{7}$_{\rm SWCX}/$\ion{O}{7}$_{\rm Total}$ 
  &
$0.24^{+0.44}_{-0.24}$
& 
$0.23^{+0.42}_{-0.23}$
  &
$0.23^{+0.44}_{-0.23}$ \\
\enddata
\tablecomments{All uncertainties and limits are reported at 90\% confidence level.}
\tablenotetext{a}{See footnotes to Table~\ref{tbl-3}.}
\end{deluxetable}

\clearpage

\end{document}